\documentclass[twocolumn,english,aps,prl,superscriptaddress]{revtex4}
\usepackage[latin9]{inputenc}
\usepackage{amsmath}
\usepackage{amssymb}
\usepackage{esint}
\usepackage{epsfig}
\usepackage{color}
\usepackage{ulem}
\usepackage{verbatim}

\date{\today}
\makeatletter
\@ifundefined{textcolor}{}
{%
 \definecolor{BLACK}{gray}{0}
 \definecolor{WHITE}{gray}{1}
 \definecolor{RED}{rgb}{1,0,0}
 \definecolor{GREEN}{rgb}{0,1,0}
 \definecolor{BLUE}{rgb}{0,0,1}
 \definecolor{DBLUE}{rgb}{0.2,0.2,0.6}
 \definecolor{CYAN}{cmyk}{1,0,0,0}
 \definecolor{MAGENTA}{cmyk}{0,1,0,0}
 \definecolor{YELLOW}{cmyk}{0,0,1,0}
}
\usepackage{babel}
\def\PT{\mathcal{PT}}
\def\P{\mathcal{P}}
\def\T{\mathcal{T}}
\def\C{\mathbb{C}}
\DeclareMathOperator{\arcsinh}{arcsinh}

\def\blgn#1\elgn{\begin{align}#1\end{align}}
\def\om{\omega}
\def\eps{\epsilon}
\def\omb{(\omega)}
\def\Sig{\Sigma}
\def\f{\frac}

\def\mcG{\mathcal G}

\newcommand{\fref}[1]{Fig.~\ref{#1}}
\newcommand{\Fref}[1]{Fig.~\ref{#1}}

\makeatother

\usepackage{babel}
\begin{document}

\title{Parity-time symmetry-breaking mechanism of dynamic Mott transitions
in dissipative systems}
\author{Vikram Tripathi}
\affiliation{Materials Science Division, Argonne National Laboratory,
	9700 S. Cass Avenue, Argonne, Illinois 60439, USA} 
\affiliation{Department of Theoretical Physics, Tata Institute of Fundamental Research, Homi Bhabha Road, Mumbai 400005, India}
\author{Alexey Galda}
\affiliation{Materials Science Division, Argonne National Laboratory,
	9700 S. Cass Avenue, Argonne, Illinois 60439, USA}
\author{Himadri Barman}
\affiliation{Department of Theoretical Physics, Tata Institute of Fundamental Research, Homi Bhabha Road, Mumbai 400005, India}
\author{Valerii M. Vinokur}
\affiliation{Materials Science Division, Argonne National Laboratory,
	9700 S. Cass Avenue, Argonne, Illinois 60439, USA}

\begin{abstract}
We describe the critical behavior of electric field-driven (dynamic) Mott insulator-to-metal transitions in dissipative Fermi and Bose systems in terms 
of non-Hermitian Hamiltonians invariant under simultaneous parity ($\P$) and time-reversal ($\T$) operations. The dynamic Mott transition is identified 
as a $\PT$ symmetry-breaking phase transition, with the Mott insulating state corresponding to the regime of unbroken $\PT$ symmetry with a real energy 
spectrum. We establish that the imaginary part of the Hamiltonian arises from the combined effects of the driving field and inherent dissipation. We derive 
the renormalization and collapse of the Mott gap at the dielectric breakdown and describe the resulting critical behavior of transport characteristics. 
The obtained critical exponent is in an excellent agreement with experimental findings.
\end{abstract}

\maketitle

Non-Hermitian $\PT$-symmetric quantum Hamiltonian models introduced in the seminal work of Bender and Boettcher~\cite{Bender} offer a 
foundation for the description of non-equilibrium steady states of dissipative quantum systems~\cite{rubinstein,graefe,prosen,bhaseen}. 
The basic property of non-Hermitian $\PT$-symmetric models is that their eigenstates exhibit a continuous $\PT$ symmetry-breaking 
phase transition when the strength of the external non-conservative driving force exceeds a certain threshold value. Below this threshold, i.e. in the 
regime of \textit{unbroken} $\PT$ symmetry, the energy eigenvalues are real, while above it the energy spectrum acquires an imaginary part. In this Letter 
we demonstrate that the imaginary part of a $\PT$-symmetric Hamiltonian arises from the combined effects of the driving field and inherent dissipation. 
This offers a perfect framework for the theoretical description of out-of-equilibrium open quantum systems and dynamic transitions that occur in them.

$\PT$-symmetric models arise across the entire non-equilibrium physics and describe optical waveguides~\cite{Kottos}, electric 
RLC circuits~\cite{Schindler}, microwave cavities~\cite{Peng} and superconducting wires~\cite{vinokur}, to name a few. Here we focus 
on a theory of electric field or current driven Mott metal-insulator transitions (MIT) as a $\PT$ symmetry-breaking phenomenon. The interest in 
non-equilibrium MIT is motivated by both the intellectual appeal of understanding dynamic instabilities in 
quantum many-body strongly correlated systems and the high technological promise of Mott systems as a platform for switching devices 
in emergent electronics~\cite{Ahn}. There have been tantalizing reports of field-driven Mott MIT in VO$_{2}$~\cite{Kim,Laad}, 
La$_{2-x}$Sr$_{x}$NiO$_{4}$~\cite{Yamanouchi}, one-dimensional Mott insulators Sr$_{2}$CuO$_{3}$/SrCuO$_{2}$~\cite{Taguchi}, and organic 
compounds~\cite{Kumai}, yet the critical behavior at the dynamic Mott transition remained unexplored. In a recent experimental 
breakthrough~\cite{Science:2015}, the current-driven Mott transition has been observed in a system of vortices pinned by a periodic array of 
proximity coupled superconducting islands.
Notably, the revealed critical behavior of the dynamic resistance near the dynamic Mott critical point appeared to belong to the liquid-gas
transition universality class. 
Here we propose the $\PT$ symmetry-breaking mechanism of dynamic MIT and find the dynamic critical behavior in full accord with the experiment.

There has been a remarkable progress in unearthing the mechanism 
of the field-driven breakdown of the Mott insulator, which was identified as the
Landau-Zener-Schwinger (LZS) process of generation 
of free particle-hole excitations by an external driving field~\cite{oka,eckstein,green}. The remaining puzzle concerns description of 
the collapse of the Mott gap at the transition. As we show below, this can be achieved by taking into account dissipation processes.
%The commonly accepted mechanism for a field-driven breakdown of a Mott insulator is Landau-Zener-Schwinger (LZS) mechanism of generation 
%of free particle-hole excitations by an external driving field~\cite{oka,eckstein,green}. 
%Yet, the above studies did not take into account dissipation processes, neither did they describe the renormalization and collapse of the 
%Mott gap at the transition. 
A recent numerical study~\cite{aron} that included dissipation still did not address the critical behavior. 
An important step towards including the dissipation effects into the picture was taken in~\cite{fukui} via the Bethe ansatz treatment of a half-filled Hubbard chain subject 
to a constant imaginary gauge field, the approach resembling the delocalization transition induced in a system of noninteracting vortices by an 
imaginary vector potential~\cite{hatano}. 
In an intriguing parallel development in high-energy physics, a numerical treatment of the Schwinger mechanism 
in Scalar Electrodynamics revealed mass renormalization due to thermalization of produced particles~\cite{gelis}.

Here we address the challenge of description of the collapse of the Mott gap at the transition. We develop a theory of the electric field-driven MIT
based on the concept of the ${\cal PT}$ symmetry breaking.
We show that it is the applied electric field which, in the presence of dissipation, generates an imaginary part of the system's Hamiltonian while 
retaining its $\PT$ symmetry. We consider fermionic and bosonic systems that undergo the transition and identify their MITs  as $\cal{PT}$ 
symmetry-breaking phase transitions. For a 
half-filled Hubbard chain we adopt the Bethe ansatz approach~\cite{fukui} and obtain the critical scaling of the Mott gap $\Delta$ with driving field $F,$ 
$\Delta \sim (F_c - F)^{1/2}$. Then we find the probability $P\sim e^{-2\gamma}$ for the LZS dielectric breakdown with the LZS tunneling 
parameter $\gamma\sim (F_c - F)^{3/2}$.
For a two-dimensional model we employ a dynamical mean field theory (DMFT) approach with an iterative perturbation theory (IPT) based impurity 
solver~\cite{hbar:nsv:ijmpb11} and find a critical scaling $\Delta \sim (F_c - F)^{0.78\pm0.03}.$ Finally, for the vortex (bosonic) Mott 
transitions driven by a current $I,$ we obtain $\Delta \sim (I_c - I)^{1/2},$ for the scaling of the spectral gap,
and $\gamma \sim (I_c - I)^{3/2}$ for the LZS tunneling parameter,
in excellent accord with recent experimental results of Ref.~\cite{Science:2015}. 

\begin{figure}
\includegraphics[width=0.9\linewidth]{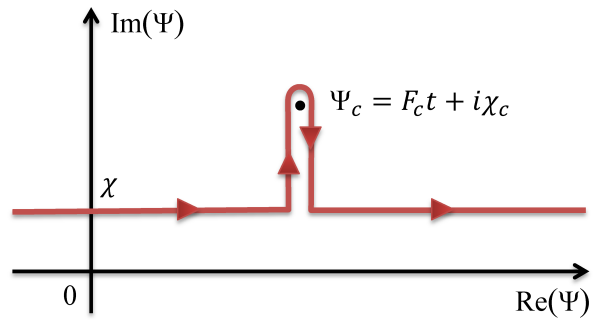} 
\caption{Schematic of the contour in the complex $\Psi$ plane for calculation of the Landau-Zener-Schwinger transition probability. 
The point $\Psi_c = F_ct + i\chi_c$ is a degeneracy point where the energy gap, ${\Delta \equiv E_1 - E_0}$, closes. Only the parts of 
the contour parallel to the imaginary axis contribute to the transition probability. For finite dissipation, the vertical part of the contour 
begins from a nonzero value of $\chi$, whose implicit dependence on the dissipation and driving field is obtained in the main text. 
%The gap is found to close as $(\chi_c - \chi)^{1/2}$ for the two examples studied here.
}
\label{fig:contour}
\end{figure}

\textit{The model--} %The existence of a background driving field promotes
%conduction in a gapped system (i) through generation
% %magnitude fixed, and (ii) by possible renormalization of the energy
%gap (mass) by the background field which then affects the LZS tunneling
%probability. 
Let $|0\rangle$ be the ground state of an interacting quantum system, $|1\rangle$ be the lowest excited state, and 
${\Delta = E_1 - E_0}$ be a spectral gap, with $E_0$ and $E_1$ being the eigenvalues for the ground and lowest excited state, 
respectively. Within the LZS framework, the electric field-induced probability for a ${|0\rangle \to |1\rangle}$ 
transition is given by the Landau-Dykhne formula~\cite{Dykhne:1962}, ${P  =|\langle0|1\rangle|^{2}\sim e^{-2\gamma}}$, where 
${\gamma = (1/\hbar)\text{Im}\int_{-\infty}^{\infty}dt\,\Delta\!\left[\Psi(t)\right]}$.
Here $\Psi$ is a time-dependent phase factor related to the driving field, $F$.  For electrons hopping along a 1D lattice under 
a constant electric field, we choose the gauge where the driving field is the time derivative of a vector potential.
We replace the integration over time by that over complex $\Psi = Ft+ i\chi(F)$, where the imaginary component $\chi$, 
shown below to be directly responsible for energy gap renormalization, originates from the effects of dissipation and 
is an odd function of $F$.  
Assuming that $\Psi(F)$ is well-behaved and that at some critical $\chi_c \equiv \chi(F_c)$ the gap completely closes, the imaginary part of the integral 
over $\Psi$ comes from the degeneracy point $i\chi_c$, see Fig.~(\ref{fig:contour}):
\begin{align}
	\gamma & \sim \frac{1}{F}\text{Re}\int_{\chi}^{\chi_c} d\chi'[E_{1}(\chi')-E_{0}(\chi')]\,.\label{eq:oka2}
\end{align}
In non-dissipative models, the Landau-Dykhne formula reduces to the Landau-Zener result, 
${\gamma \sim \Delta^{2}/vF}$, where ${v = |d\Delta/dt|/F}$ is the speed of convergence of the two levels as $\Psi$ is varied, 
and is assumed to be a constant (i.e. independent of $\Delta$) in the usual Landau-Zener analysis. Writing ${\gamma = F_{\text{th}}/F}$, 
we identify the threshold field for the LZS tunneling,  $F_{\text{th}} = \Delta^{2}/v$. The same relation has been obtained in 
Ref.~\cite{oka} for the half-filled Hubbard chain.

To obtain the dependence of the spectral gap on the imaginary gauge field $\chi$ we solve an auxiliary problem where we set the real part of $\Psi$ to zero. 
The following analysis brings out a key point of this work, namely the fact that a finite imaginary component of the vector potential $\Psi$ naturally 
appears in dissipative quantum systems subject to a driving electric field. Moreover, the Hamiltonian describing such a system must be $\PT$-symmetric, 
with the eigenstates breaking $\PT$ symmetry at large values of the drive. Let us consider a Legendre 
transformed Hamiltonian~\cite{antal,cardy}:
\begin{align}
	H' & =H-i\lambda J \label{eq:Lagrange}\,,
\end{align}
where $H$ is Hamiltonian of the equilibrium system with a gapped energy spectrum, $J$ is the current operator commuting with $H$, and $\lambda \in \C$ is 
a Lagrange multiplier. 
%We expect the ground state of $H'$ to depend analytically on $\lambda$, meaning that, 
For sufficiently small $\lambda$, the spectral gap in $H$ implies that the expectation value of $J$ vanishes. 
In the opposite limit of large $\lambda$, the eigenfunctions of $H'$ are essentially those of $J$, and the system appears in a gapless phase with a finite 
steady current $I$. The phase transition between the equilibrium and the finite current-carrying states takes place at some critical value ${|\lambda| = \lambda_c}$, 
and the corresponding value of the Lagrange multiplier is related to the current by ${I = \langle J(\lambda)\rangle}$. Since $H$ and $J$ are both 
Hermitian operators, for purely real $\lambda$ the eigenvalues of $H'$ are real only when ${\langle J(\lambda)\rangle = 0}$, and complex when 
${\langle J(\lambda)\rangle \neq 0}$. Real $\lambda$ in Eq.~(\ref{eq:Lagrange}) corresponds to an imaginary vector potential, and, most importantly, 
guarantees inherent $\PT$ symmetry of $H'$. From the viewpoint of $\PT$-symmetric quantum mechanics, the parametric region ${|\lambda| < \lambda_c}$ 
corresponds to a regime where the eigenstates preserve $\cal{PT}$ symmetry resulting in a real spectrum for $H'$ and zero steady current, 
while for ${|\lambda| > \lambda_c}$ the spectrum of $H'$ acquires an imaginary component, and the energy gap closes leading to a finite 
steady current $I$, marking the transition into the phase of broken $\PT$ symmetry. 

To establish that real $\lambda$ describes a non-equilibrium steady state of a dissipative driven system, we consider the dynamic 
equation for the density matrix $\rho$: 
\begin{align}
	\frac{d\rho}{dt} & =-i[H,\rho] - \lambda\bigl(\{J,\rho\} - 2\text{tr}(\rho J)\rho\bigr).\label{eq:neumann}
\end{align}
For pure states, Eq.\,(\ref{eq:neumann}) reduces to the Schr\"{o}dinger equation~\cite{brody} with the non-Hermitian Hamiltonian from Eq.~(\ref{eq:Lagrange}). 
The formal solution to Eq.\,(\ref{eq:neumann}) reads~\cite{brody} 
\begin{equation}
	\rho(t) =\frac{e^{-i(H-i\lambda J)t}\rho(0)e^{i(H + i\lambda J)t}} {\text{tr}\left(e^{-i(H - i\lambda J)t}\rho(0)e^{i(H + i\lambda J)t}\right)}\,,\label{eq:soln-neumann}
\end{equation}
where $\rho(0)$ is the density matrix corresponding to the initial state. Formula\,(\ref{eq:soln-neumann}) generalizes the unitary evolution 
(which corresponds to ${\lambda = 0}$) onto the dissipative case (${\lambda \neq 0}$) and preserves the norm, ${\text{tr}\rho(t) = 1}$ so that ${d\,\text{tr}(\rho(t))/dt = 0}$.
The expectation value of any physical observable $A$ is 
${\langle A(t)\rangle = \text{tr}\left( \rho(t)A\right)}$, so the expectation value of the current operator is
\begin{equation}
	\frac{d\langle J\rangle}{dt} =-2\lambda\left( \langle J^{2}\rangle-\langle J\rangle^{2}\right)\,.\label{eq:current-decay}
\end{equation}
Therefore, if $\lambda$ is real, the system relaxes to a state with constant current 
since ${\langle J^{2}\rangle - \langle J\rangle^{2}\geq 0}.$
This, in turn, means that the transition between the equilibrium zero-current and steady-current states caused by varying the Lagrange 
multiplier $\lambda$ mirrors the electric field-driven dynamic Mott transition: below some critical field $F_c$ 
(corresponding to ${\lambda_c}$) the spectral gap is finite and ${I = 0}$, while at larger applied fields a finite current flows with a
magnitude that increases monotonically with $F$ (or $\lambda$).
%We have thus shown that the mechanism of transition between the equilibrium zero-current and steady-current states by varying the Lagrange 
%multiplier $\lambda$ is consistent with the electric field-driven dynamic Mott transition: below some critical field $F_c$ 
%(corresponding to ${\lambda_c}$) the spectral gap is finite and ${I = 0}$, while at larger applied fields a finite current flows with 
%magnitude that increases monotonically with $F$ ($\lambda$).

Now we are equipped to complete the derivation of the electric field-induced suppression of the Landau-Zener parameter $\gamma(\chi)$, and its ultimate vanishing 
at ${\chi = \chi_c}$. Provided that ${\Delta(\chi)}$ is a continuous function of $\chi$, the energy 
gap vanishes smoothly as $\chi \to \chi_c.$ The fermionic and bosonic Mott insulator systems we study all exhibit a power-law collapse for 
the spectral gap, ${\Delta(\chi)\sim (\chi_c -\chi)^\alpha}$,
with $\alpha=0.5$ and $\alpha = 0.78\pm 0.03$ respectively for the 1D and 2D half-filled Hubbard models, and $\alpha=0.5$ for the (bosonic) vortex Mott insulator. 
This presents one of the main results of this Letter. 
Similar field-induced gap renormalizations have been reported in studies of other non-Hermitian models, with $\alpha=0.5$
for the non-Hermitian $XXZ$ chain~\cite{albertini} experiencing an Ising transition. Dynamic phase transition of the same 
universality class has also been predicted for a quasi-1D superconducting wire subject to imaginary vector potential destroying superconducting 
fluctuations beyond a certain threshold~\cite{rubinstein,vinokur}.

\textit{Fermionic dynamic Mott transition--} To derive critical behavior of the dynamic Mott transition, we write 
the Hamiltonian of the half-filled fermionic Hubbard chain in presence of the imaginary component $\chi$ of vector potential as 
\begin{equation}
	H = -t\sum_{\langle ij\rangle,\sigma} [e^{\chi}c_{i\sigma}^{\dagger}c_{j\sigma} + e^{-\chi}c_{j\sigma}^{\dagger}c_{i\sigma}] + 
	U\sum_{i}n_{i\uparrow}n_{i\downarrow},\label{Hubbard-H}
\end{equation}
where $t$ is the hopping amplitude, ${U > 0}$ is the on-site repulsive Coulomb interaction strength, and after rescaling the bandwidth to the zero field case,
we identify $\tanh(\chi)$ as the Lagrange multiplier $\lambda$ from Eq.\,(\ref{eq:Lagrange}).

Starting from the Bethe ansatz solution~\cite{fukui} of Eq.\ref{Hubbard-H}, one obtains the following expression for the Mott gap~\cite{lieb} 
on the insulating side of the transition:
\begin{align}
	\Delta(b) & = 4t\left[u - \cosh(b) + \int_{-\infty}^{\infty}\frac{d\omega}{2\pi}\frac{J_{1}(\omega)e^{\omega\sinh(b)}}{\omega(1 + 2^{2u|\omega|})}\right]\,,\label{delta}
\end{align}
where $u \equiv U/4t$, $J_1$ is the Bessel function of order $1$, and ${b \leq b_c}$ is the parameter controlling the integration contour for the spin distribution 
function in the complex plane (see Ref.~\cite{fukui}), such that ${b_c \equiv \arcsinh(u)}$ defines the point of Mott transition, ${\Delta(b_c) = 0}$.
We find that as $\chi\to\chi_c,$ ${\chi_c - \chi \approx C_1(b - b_c)^2},$ which in turn leads to ${\Delta(\chi) \approx C_2 \sqrt{\chi_c - \chi}}$, 
where $C_{1,2}$ are constants. Assuming that $F(\chi)$ is a well-behaved function at the threshold, we have 
${\Delta(F) \sim \sqrt{F_c - F}}$, and we get
\begin{align}
\gamma\sim (F_c - F)^{3/2}\, \label{scaling}
\end{align}
%Expanding ${\chi = f(F\rho)}$ near $\chi_c$, it follows that 
%For driving fields above critical, such that ${\chi > \chi_{\text{cr}}}$, the energy spectrum becomes complex heralding breaking down of $\cal{PT}$ symmetry.
for the scaling of the LZS tunneling parameter.

\begin{comment}
Now we consider a generalization of the above model to two dimensions; our model is now a 2D half-filled square lattice with
nearest-neighbor hopping and a nonequilibrium drive along the $x$-direction:
\begin{align}
 H & = \sum_{\mathbf{k},\sigma}[-2(\cos(k_x)+\cos(k_y)) - i\lambda\sin(k_x)-\mu]c^{\dagger}_{\mathbf{k}\sigma}c_{\mathbf{k}\sigma} \nonumber \\
   & \qquad + U\sum_{i}n_{i\uparrow}n_{i\downarrow},
 \label{eq:hubbard_2D}
\end{align}
where $\lambda\in[0,2]$ is the (real) Lagrange multiplier introducing the current constraint as usual. We employ a DMFT technique
based on an iterative perturbation expansion (see S.I. for details) for the calculation of the spectral function for single-particle
excitations from which we extract the dependence of the Mott gap on $\lambda.$ Fig.~\ref{fig:mott-gap} shows the behavior of the Mott gap
for $U=30$ upon approaching the transition from the insulating and metallic directions. A power-law fit to the Mott gap yields an exponent $0.52$
which to numerical accuracy is identical to the exponent in the one-dimensional case. 
\end{comment}

\begin{figure}
\includegraphics[width=0.95\linewidth,clip]{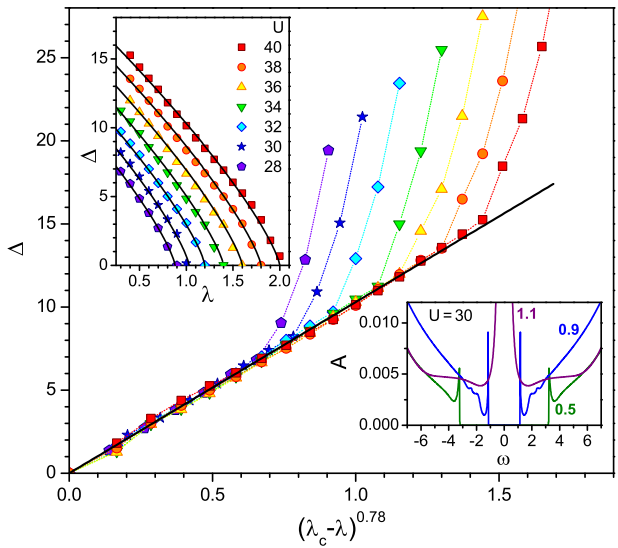}
\caption{(Color online) 
Universal scaling of the Mott gap $\Delta$ as a function of the drive $\lambda$ near the $\PT$ symmetry breaking points 
$\lambda_c$ for different values of the Coulomb repulsion $U.$
The upper inset shows the evolutions of the Mott gaps with 
increasing $\lambda$ for various $U,$ and the solid lines are fits to power-laws of the form $\Delta =  C(\lambda_c - \lambda)^{0.78}.$
The main figure shows the same data 
plotted as function of $(\lambda_c - \lambda)^{0.78}$ 
(the legends for the inset and the main panel are the same). 
One sees a remarkable linear collapse of the data persisting over large
region of $\lambda$. 
The lower inset shows the $\lambda$-dependence of the spectral 
function $A(\omega)$ for single-particle excitations for $U=30.$ As $\lambda$ increases, the spectral gap gradually narrows, and for ${\lambda = 1.1}$
a quasiparticle band is evident signifying a (bad) metallic phase.}\label{fig:mott-gap}
\end{figure}

Now we consider a 2D Hubbard model, a half-filled square lattice  with nearest-neighbor hopping and non-equilibrium drive 
along the $x$-direction:
\begin{align}
	H & \!=\! \sum_{\mathbf{k},\sigma}\!\left\{-t\left[2(\cos(k_x) \!+\! \cos(k_y)) \!-\! i\lambda\sin(k_x)\right] \!-\! 
	\mu\right\}c^{\dagger}_{\mathbf{k}\sigma}c_{\mathbf{k}\sigma} \nonumber \\
	& \qquad + U\sum_{i}n_{i\uparrow}n_{i\downarrow}\,,\label{eq:hubbard_2D}
\end{align}
where ${\lambda \in [0,2]}$ is the real Lagrange multiplier introducing the current constraint. In what follows we calculate the spectral functions by employing a second-order perturbation theory approximation, 
namely the IPT within the DMFT framework~\cite{hbar:nsv:ijmpb11} (see Appendix for details). For a sufficiently small driving 
field $\lambda$ and large interaction strength $U,$ a Mott gap is formed at the Fermi level (${\omega = 0}$), as seen from 
the spectral function in the lower inset of Fig.~\ref{fig:mott-gap}. Gradual increase of $\lambda$ diminishes the gap and eventually, for 
${\lambda \geq \lambda_c}$, a quasi-particle peak appears at the Fermi level by closing the gap and, hence, signifying an insulator-to-metal 
transition. In the upper inset of Fig.~\ref{fig:mott-gap}, we present the Mott gap $\Delta$ (extracted from the calculated spectral function) for different
values of $U$ as a function of $\lambda$. Plotting the same data as function of $(\lambda - \lambda_c(U))^{\alpha}$ we find them to collapse to a straight line for $\alpha=0.78$, see the main panel. From such 
collapse of the data we infer a universal (i.e. independent of $U$) power-law behavior with a critical exponent $0.78\pm 0.03.$
This exponent, larger than $0.5$ we obtained above for 1D, lies closer to the exponent $1$ reported 
in mean-field studies of the equilibrium 2D Mott transition~\cite{brinkman,florens}.

%% FIGURE : 

\textit{Bosonic dynamic Mott transition-- } Let us now consider a bosonic system described by a non-relativistic Landau-Ginzburg-Wilson (LGW) field theory. As an example, we take
vortices in an array of traps in 
magnetic fields corresponding to integer fillings $f_c.$ Vortex Mott insulator state has been predicted in~\cite{NelsVin:1993} 
and observed in numerous experiments, see Refs.\,\cite{Harada:1996,zeldov} and references therein. 
Moreover, we can now justly believe that the vortex Mott insulator have been seen, albeit not recognized, in numerous studies of vortex matching effects in nanopatterned superconductors~\cite{pratap,baturina2010}.
The critical scaling behavior as a function of the driving 
current, temperature, and departure from commensuration ${|f - f_{c}|}$ at the dynamic vortex Mott transition has been observed in square proximity 
superconducting arrays\,\cite{Science:2015}. 

\iffalse
An explanation for the current-driven transition at integer fillings $f_{c} = 1,2,\ldots$ was provided in terms of a simple model 
of a current-driven resistively-shunted Josephson junction. For a junction with a critical current $I_{c},$ the effective barrier for phase 
delocalization due to Landau-Zener tunneling scales as $(I_{c}-I)^{3/2}$ which is consistent with the observed collapse of the transconductance 
data as a function of $|I-I_{c}|/|f-f_{c}|^{2/3}$ near integer fillings. Below we present an alternate derivation of this result which brings
out the $PT$ symmetry-breaking mechanism of this transition clearly.
\fi

Near ${f_{c} = 1}$, 
%careful Monte-Carlo studies\,\cite{hebert} show that
the vortex Mott transition is mean-field like, and its dynamics can be described by a non-relativistic LGW 
effective action in Euclidean time, 
\begin{equation}
	S \!=\! \int\! d^{2}x\, d\tau\!\left[\psi^{\dagger}\frac{\partial}{\partial\tau}\psi  \!+\! D|\nabla\psi|^{2} 
	\!+\! m^{2}|\psi|^{2} \!+\! u|\psi|^{4}\right]\,.\label{eq:LG-model}
\end{equation}
Here $\psi$ is the vortex field, $D$ is the vortex stiffness, $m$ and $u$ are, respectively, the mass and 
interaction parameters that govern the mean-field transition, where the `superfluid' phase of vortices corresponds to 
${m^{2} < 0}$. The applied electric current, $I$, exerts Magnus force on magnetic vortices. We incorporate 
the current into the vector potential: ${A_{x} = It,\, A_{y} = 0}$. 
We consider vortex motion in a dissipative 
region surrounded by a superconducting shell. This enables us to impose the simple boundary condition ${\psi = 0}$ 
outside the dissipative region. The motion in the dissipative environment is overdamped, thus the time evolution is governed entirely  by 
Brownian processes and we can neglect Berry phase effects [first term in Eq.\,(\ref{eq:LG-model})]. In real time, 
the equation of motion is 
\begin{equation}
	\frac{\partial\psi}{\partial t} + \nu\frac{\delta H}{\delta\psi^{*}} = 0\,,\label{eq:eq-motion}
\end{equation}
where ${H = \int d^{2}x\,\left[D|\nabla\psi|^{2} + m^{2}|\psi|^{2} + u|\psi|^{4}\right]}$ is the Hamiltonian corresponding 
to Eq.\,(\ref{eq:LG-model}), and $\nu$ represents viscous damping of the vortex motion and is proportional 
to the normal resistance of the superconductor just above superconducting transition temperature $T_{\mathrm c}$. 
Following Refs.\,\cite{rubinstein,vinokur}, we ignore the non-linear term in the vicinity of the transition and after a 
straightforward calculation arrive at the following expression for the distance between the two lowest-energy levels on the 
vortex Mott-insulating side ($I \leq I_c$):
\begin{equation}
	E_{1} - E_{0} \approx 2E_{T} \sqrt{\eta(1 - I^{2}/I_c^{2})}\,,
 \label{eq:hopf}
\end{equation}
where for a size $L$ of the normal region where the vortex is confined, ${E_{T} = D/L^{2}}$ is the Thouless energy, 
${\eta \approx (\pi^{2}/\sqrt{2})(I_c L/E_{T}\rho)}.$

Interpreting the vortex excitation gap as the Mott gap, we obtain the square-root scaling behavior near the transition, and from the 
Landau-Dykhne formula find the barrier for vortex thermal activation
%, to which Eq.\,\eqref{eq:oka1} maps, 
to scale as
\begin{equation}
	\gamma \sim (I_c - I)^{3/2}\,.
\end{equation}

%Considering that the current
%in the vortex system effectively acts as an electric field in a corresponding
%half-filled electronic system, we expect a Mott transition when the
%energy $I_{\text{th}}a$ gained in a nearest-neighbor hop in the presence
%of the external field becomes comparable with the effective ``Hubbard''
%repulsion $|f-f_{c}|.$ 
This result is in excellent agreement with the experimental findings of Ref.~\cite{Science:2015} demonstrating scaling of the dynamic 
resistance near the vortex Mott transition as a function of ${|I_c - I|^{3/2}/|f - f_{c}|}$.

\textit{Discussion--} To summarize, we investigated Mott transitions in  fermionic half-filled Hubbard models in one and two dimensions and in bosonic vortex lattice system near integer fillings. 
We showed that non-equilibrium steady states of such systems are described as eigenstates of non-Hermitian Hamiltonians endowed with $\cal{PT}$ symmetry. 
The field-driven Mott transition is identified as a $\cal{PT}$ symmetry-breaking phenomenon. We related the driving electric field and the dissipation 
parameter to the non-Hermitian gauge fields governing the $\cal{PT}$ symmetry-breaking phase transition. While the mechanism of Mott transitions in 
these dissipative systems, Landau-Zener-Schwinger tunneling, is also shared with non-dissipative and even non-interacting quantum systems, the key qualitative 
difference in the dissipative case lies in the field-induced renormalization of the excitation gap. For the 1D Hubbard chain and the bosonic 2D system, 
we find that the spectral gap $\Delta$ and the LZS tunneling factor $\gamma$ respectively scale as $\Delta\sim(F_c - F)^{1/2}$ and 
${\gamma\sim(F_{c} - F)^{3/2}}$ as a function of the driving field $F$. This behavior 
is in accord with the current vs. magnetic field scaling recently observed in the vortex Mott transition in nanopatterned 
superconductors~\cite{Science:2015}. For the 2D Hubbard model, we perform a DMFT analysis based on an IPT approximation scheme and obtain 
scaling $\Delta\sim(F_c - F)^{0.78}.$ 
%and compare that to the Mott criticalities brough about purely through tuning the $U/t$ ratio or the filling.
One of the remaining open problems to address 
is the microscopic derivation of the effective non-Hermitian Hubbard models starting from the driven Hermitian system coupled to a bath. Note, finally, that
vortex insulator-metal transitions have also been suggested in quantum Hall systems~\cite{shahar}, which could provide a new platform for exploring
dynamic Mott transitions.

Another direction to take, is the
study into the role of disorder in the  $\cal{PT}$ symmetry-breaking transitions are needed. The past researches of effects of disorder, such as vortex 
delocalization driven by an imaginary vector potential~\cite{hatano} and level-statistics of zero-dimensional fermionic systems~\cite{efetov}, are restricted 
to non-interacting systems. The incoherent hopping transport in disordered insulators in moderately strong electric fields, which is known to be of the 
directed-percolation type driven by an imaginary field~\cite{obukhov,cardy-sugar}, %so that they apparently belong in the class of non-Hermitian systems with  $\cal{PT}$ symmetry,
will be the subject of forthcoming publication. Our work thus paves the way for the application of $\cal{PT}$ symmetry concepts as a general mechanism 
of the dynamic phase transitions in strongly correlated quantum systems.

\begin{acknowledgments}
We thank T.\,I.\,Baturina for critical reading of the manuscript and many valuable suggestions, and D.\,Dhar and T.\,V.\,Ramakrishnan for illuminating discussions. 
The work is supported by the U.S. Department of Energy, Office of Science, Materials Sciences and Engineering Division,
(V.T. is partially supported by Materials Theory Institute at ANL),
by the University of Chicago Center in Delhi, 
and DST (India) Swarnajayanti grant (no. DST/SJF/PSA-0212012-13). 
H.B. is grateful for support from DAE (India) and computational 
resources from the Department of Theoretical Physics, TIFR.
\end{acknowledgments}

\appendix

\section{Dynamical mean-field theory}

We numerically study the dissipative fermionic Hubbard model in a two-dimensional tight-binding lattice using the dynamical mean-field 
theory (DMFT), which maps the correlated model onto an effective interacting impurity model, where the impurity is self-consistently coupled 
to a non-interacting fermionic bath~\cite{georges:kotliar:krauth:rozenberg:rmp96}. We solve the effective impurity model using the iterated 
perturbation theory (IPT), which has been extensively applied to various correlated models in the context of DMFT. At half-filling, the IPT 
self-energy is merely the second order perturbation expansion in $U$ around the Hartree limit:
\begin{equation}
\Sig^{\text{IPT}}\omb=\f{U}{2}+\Sig_2\omb;\nonumber
\end{equation}
\begin{equation}
\Sigma_2(\om) = \lim_{i\om\to\om^+}\frac{U^2}{\beta^2}\sum_{m,p}{\mcG}(i\om + i\nu_m)
{\mcG}(i\om_p + i\nu_m){\mcG}(i\om_p)\,,
\end{equation}
where $\mcG$ is the bath propagator hybridized with the impurity in the non-interacting case (${U = 0}$). Here we employ an improved 
version of the direct real-frequency implementation of the method, which bypasses the rigor of numerical analytical continuation and handles 
the divergence in self-energy in the insulating regime by expressing it through an analytical expression~\cite{hbar:nsv:ijmpb11}.
%===================
\section{Results}
\subsection{Non-interacting density of states}
\begin{figure}
	\includegraphics[width=0.8\linewidth]{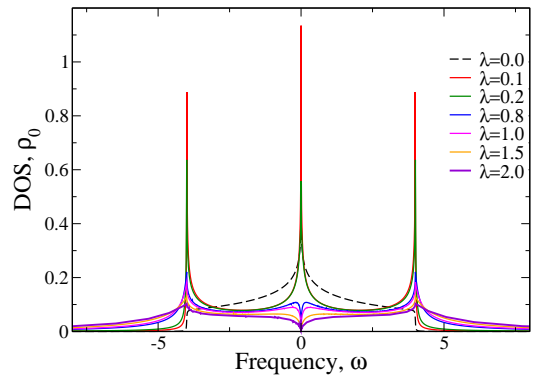}
	\caption{DOS for the driven two-dimensional Hubbard model. $\lambda>0$ gives rise to additional van Hove singularities near the boundaries 
	signifying a presence of 1D-drive. At ${\lambda\ge 0.8}$, the singularity at ${\omega = 0}$ disappears reflecting a possible 2D to 1D 
	crossover.}\label{fig:dos:var:lambda}
\end{figure}

%-------------------------------------------------------------
We first find the non-interacting lattice density of states (DOS) from the energy dispersion for Eq.~(19) in the main 
text: ${\eps(k_x,k_y) = -2t(\cos k_x + \cos k_y) - i\lambda\sin k_x}$. We sum over all momentum points $k_x$ and $k_y$ in the first 
Brillouin zone in order to get the DOS: 
\begin{equation}
\rho_0(\omega) = \frac{1}{(2\pi)^2}\sum_{k_x,k_y}\frac{1}{\omega^+ - \eps(k_x,k_y)}\,.   
\end{equation}
In our computation presented in Fig.~\ref{fig:dos:var:lambda} we set ${t = 1}$, which serves as the unit of all energy parameters 
(e.g. $U$) in our final DMFT results. At ${\lambda = 0}$, the DOS shows a usual square lattice tight-binding DOS with a presence of  van 
Hove singularity (VHS) at the center (${\omega = 0}$) and the energy band remains bounded between $-4t$ and $4t$. As soon as $\lambda$ is turned on, 
the DOS boundaries get extended and subsidiary VHS arises near the boundaries due to presence of 1D drive along the $x$-direction. At ${\lambda \ge 0.8}$, 
the VHS at the center disappears, signifying a possible crossover of the DOS from 2D to 1D in nature. Confirmation of such crossover and its effect 
is left for our future work. 
%% 1st figure
%-------------------------------------------------------------
\subsection{Spectral functions of the interacting Hamiltonian}
Now we look at the single particle spectral functions of our model by performing a Hilbert transform over the non-interacting DOS:
\begin{equation}
A(\omega) = \int_{-\infty}^\infty d\om'\, \frac{\rho_0(\om')}{\gamma(\om) - \om'}\,,
\end{equation}
where $\gamma(\om)\equiv\om^+ + \mu-\Sig^{\text{IPT}}(\om)$.

\begin{figure}
\includegraphics[width=0.8\linewidth,clip]{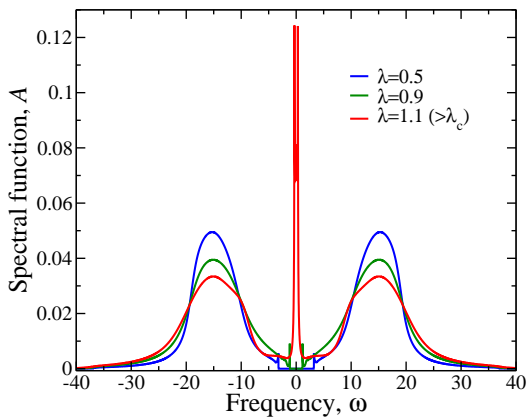}
%\includegraphics[height=0.9\linewidth,clip]{FIGS/Sp_density_lambda_increasing_U30.eps}
%\label{fig:spf:inreasing:lambda}
\caption{The spectral function $A(\omega)$ for different drive parameters $\lambda$ and fixed Coulomb repulsion $U=30.$ Increasing $\lambda$ from 
the insulator side of the transition reduces the Mott gap, with the gap completely closed at ${\lambda \geq \lambda_{c} \simeq 1.0}$.
The collapse of the Mott gap at $\omega=0$ is shown in more detail in the lower inset of Fig. 2.}\label{fig:spf:various:lambda}
\end{figure}

In the non-driven case, at small $U$, a metallic spectral function occurs with a quasiparticle resonance at the Fermi level. 
As $U$ is increased, the width of the resonance shrinks and finally disappears when $U$ exceeds a critical value $U_{c}$. Similarly, 
in the electric field-driven regime, Fig.~\ref{fig:spf:various:lambda} shows that increasing $\lambda$ diminishes the Mott gap, 
ultimately closing it completely. We remark here that for very large values of $U\gtrsim40,$ the Mott gap does not close even at $\lambda=2.$
%at values ${\lambda \geq \lambda_{c} \simeq 1.0}$.

\iffalse
%On the other hand, starting from an insulating phase, decreasing $U$ reduces the Mott gap, and ends up with a metallic 
spectral function as the interaction strength becomes lower than a critical value. However, this critical value, which is 
usually dubbed as $U_{c1}$, differs from the earlier one: ${U_{c1} < U_{c2}}$. This reflects an hysteresis, which arises due to 
presence of two solutions (metallic and insulating) of the DMFT equations in the region between $U_{c1}$ and $U_{c2}$, known 
as the coexistence regime~\cite{hbar:nsv:ijmpb11}. The coexistence regime has also been found using DMFT with other quantum 
impurity solvers~\cite{kim:choi:jkps14} and its cluster version~\cite{balzer:etal:epl09}.

%Similarly, the hysteresis is suggested by calculations in the electric field-driven regime. However, presence of 
finite $\lambda$ lets the metal-to-insulator transition happen at higher critical value, i.e. ${U_{c2} (\lambda > 0) > U_{c2}(\lambda = 0)}$. 
In others words, increasing $\lambda$ diminishes the Mott gap, and at a value ${\lambda > \lambda_{c2} \simeq 1.0}$, the gap closes. 
The presence of coexistence regime leads to hysteresis on the $\lambda$-axis as well, i.e. by decreasing $\lambda$ from a metallic regime, 
a Mott gap opens up at $\lambda$ below a different transition point ${\lambda_{c1} \simeq 0.6}$ (See \fref{fig:spf:various:lambda}).
\fi

%% 2nd figure 
%%
\begin{comment}
%--------------------------------------------------
\subsection{Closing of Mott gap: critical exponent}
Finally, we investigate the critical behavior of the transition from a gapped (Mott insulator) to a gapless (metal) phase, as a function 
of $\lambda$ for various interaction strengths. \Fref{fig:gap:closing:powerlaw:fit} shows that each curve universally follow the power-law 
${\Delta(\lambda) = C(\lambda_{c} - \lambda)^{\nu}}$, $C$ being a constant  (dashed lines show the fits). Our best fit 
shows ${C = 10}$ and ${\nu = 0.75}$, which are independent of $U$. For $U=$ 28.0, 30.0, 32.0, 34.0, 36.0, and 38.0 the obtained $\lambda_{c}$ are 
0.88, 1.03, 1.20, 1.40, 1.60, 1.80, respectively.

%% 3rd figure
\begin{figure}
\includegraphics[width=0.8\linewidth,clip]{power_law_fit_gap_closing.eps}
\caption{Critical behavior of the Mott gap $\Delta$ as the drive parameter $\lambda$ approaches the insulator-to-metal 
transition point $\lambda_{c}$. The dashed lines show universal (independent of $U$) fits: ${\Delta(\lambda) = C(\lambda_{c} - \lambda)^{\nu}}$ 
with ${C = 10}$ and ${\nu = 0.75}$.}\label{fig:gap:closing:powerlaw:fit}
\end{figure}

To do the fit we use the function ${C(\lambda_{c} - \lambda)^\nu}$, where we vary $C$ and $\nu$ to find the universal values that best fit the data for all $U$'s. 
\end{comment}

\end{document}